\documentclass[11pt]{amsart}
\usepackage{enumerate}
\usepackage{graphicx}
\usepackage{amsfonts}
\usepackage{graphicx}
\usepackage{amsmath}
\usepackage{amssymb}
\usepackage{amscd}

\newcommand{\ts}{\hspace{0.5pt}}

\newcommand{\RR}{\mathbb{R}\ts}

\newcommand{\vL}{\varLambda}

\newcommand{\Lperp}{\widetilde{L}^\perp}
\newcommand{\Lnull}{L^\circ}

\newcommand{\oplam}{\mbox{\Large $\curlywedge$ }   }
\newcommand{\gammahat}{\widehat{\gamma_\vL}}

\newcommand{\Hm}[1]{\leavevmode{\marginpar{\tiny%
$\hbox to 0mm{\hspace*{-0.5mm}$\leftarrow$\hss}%
\vcenter{\vrule depth 0.1mm height 0.1mm width \the\marginparwidth}%
\hbox to
0mm{\hss$\rightarrow$\hspace*{-0.5mm}}$\\\relax\raggedright #1}}}

\newtheorem{theorem}{Theorem}

\usepackage{rotating}

\begin{document}

\title{Aperiodic order and pure point diffraction}

\author{Daniel Lenz}
\address{Fakult\"at f\"ur Mathematik, TU Chemnitz,
09107 Chemnitz, Germany\\
Current address: Department of Mathematics, Rice University, P. O. Box 1892, Houston, TX 77251, USA\\\vspace{6pt}
}
\email{dlenz@mathematik.tu-chemnitz.de}
\urladdr{http://www.tu-chemnitz.de/mathematik/analysis/dlenz}

\maketitle

\begin{abstract} 
  We give a leisurely introduction into mathematical diffraction theory with a
  focus on pure point diffraction.  In particular, we discuss various
  characterisations of pure point diffraction and common models arising from
  cut and project schemes. We finish with a list of open problems.
\end{abstract}

\section{Introduction}
Quasicrystals were discovered by their unusual diffraction properties
\cite{SBGC}. Subsequently, quite some mathematical effort has been devoted to
diffraction theory of aperiodically ordered models (see corresponding parts
in \cite{BMbook,Mbook,Sen,Tre} and references therein). In fact, there is
substantially more work done than could sensibly be reviewed here.  For this
reason we will focus below on pure point diffraction and model sets. These
topics seem to be both particularly relevant and conceptually fairly well
understood.  We note in passing that model sets were introduced by Meyer
\cite{Mey} quite before the dawn of quasicrystals.  They became a standard in
the physical literature.

Recent results give a  diffraction theory for measures on locally compact abelian groups
\cite{BL1,BL2,BM,LR}. Here, however, we will restrict our attention  to point sets  in Euclidean space in order
to keep the exposition as simple as possible.  
The much studied topic of mixed
spectra and random systems, see e.g. \cite{BFG,BH1,BS,MPS,R} and references
therein, will only appear in the last section on problems.

For earlier survey type articles on mathematical diffraction we refer to \cite{Lag2,BMRS}.  A detailed
introduction into mathematical diffraction theory is given in the lecture
notes   \cite{Baa}.  Another  somewhat introductory text from the point of view of
dynamical systems can be found in \cite{Len2}.

The article is organised as follows: In Section \ref{Framework} we introduce
the framework of mathematical diffraction theory and define and discuss its
key quantity, the diffraction measure. Section \ref{Pure} is concerned with
characterisation of pure point diffraction.  In Section \ref{Cut} we discuss cut and project schemes and  model sets.  
Finally, in the last section we discuss some open problems.

\section{The framework of mathematical diffraction theory}\label{Framework}
Diffraction theory for crystals has a long history.  Discussions can be found
in many textbooks, e.g. \cite{Cow}.  The first systematic treatment of
mathematical diffraction theory for aperiodic order is due to Hof
\cite{Hof1,Hof2}.  Here, we discuss the basic setup.  The mathematical key
quantity turns out to be a measure, the so called diffraction measure. It
models the outcome of a diffraction experiment.

\smallskip

We will model the solid in question by a subset $\varLambda$ of Euclidean
space $\RR^N$. We will assume that $\varLambda$ is distributed in a regular
way with points being not too close and not too far. More precisely, we will
assume that $\varLambda$ is Delone, i.e.
\begin{itemize}
\item there exists an $r>0$ such that different points of $\varLambda$ have distance at least $2 r$ ("$\varLambda$ is uniformly discrete"), 
\item there exists an $R>0$ such that no point of Euclidean space has distance
  bigger than $R$ from $\varLambda$ ("$\varLambda$ is relatively dense").
\end{itemize}
Thus, clearly, $\varLambda$ has infinitely many points.  In order to
understand the diffraction setup for the infinite $\varLambda$ it is helpful
to first consider the case of a finite set $F$ of scatterers.  In this case
the intensity is the function given by
$$ I_F (\xi) = \left| \sum_{x\in F} \exp (-i x \xi) \right|^2 = \sum_{x,y\in F} \exp( - i (x-y) \xi).$$
The analogous expression for the infinite $\varLambda$ reads
$$
\sum_{x,y\in \varLambda} \exp( -i (x-y) \xi).$$
This sum is heavily
divergent, as there are infinitely many terms of modulus $1$. This is not a
mathematical problem only. There is a physical reason behind this divergence:
The intensity of the infinite solid is infinite. The correct object to be
considered is the normalised intensity per unit volume.  It is defined as
\begin{equation}\label{defI} I_\varLambda = \lim_{n\to \infty} \frac{1}{|B_n|} I_{\varLambda\cap B_n }.
\end{equation}
Here, $B_n$ denotes the ball around the origin with radius $n$ and the modulus
denotes the volume.  Note that $\varLambda \cap B_n$ is finite by our
assumption on $\varLambda$ and, hence, $I_{\varLambda \cap B_n}$ is a well
defined function. There are issues in \eqref{defI} in which sense the limit is
meant (it is the vague sense) and whether it exists at all (we may have to
pass to a subsequence). However, we will skip these somewhat technical
details.
 
The quantity $I_\varLambda$ is known as diffraction measure. It describes the
outcome of a diffraction experiment.  The values $\xi$ with $I_\varLambda
(\{\xi\}) >0$ are called Bragg peaks. The value of $I_\varLambda (\{\xi\})$ is
called intensity of the Bragg peak.

Now, that we have defined $I_\varLambda$ we may wonder how to calculate it. To
this end let us expand the definition of $I_\varLambda$. This gives
\begin{eqnarray*}
I_\varLambda = \lim_{n\to \infty} \frac{1}{|B_n|} I_{\varLambda\cap B_n } &=&\lim_{n\to \infty} \frac{1}{|B_n|} \sum_{x,y\in \varLambda\cap B_n} \exp ( -i (x-y) \xi)\\
&=& \lim_{n\to \infty} F (  \frac{1}{|B_n|}  \sum_{x,y \in \varLambda\cap B_n}   \delta_{x-y})\\
&=&           F (   \lim_{n\to \infty} \frac{1}{|B_n|}  \sum_{x,y \in \varLambda\cap B_n}   \delta_{x-y}). 
\end{eqnarray*}
Here, $\delta_x$ denotes the unit point mass at $x$ and $F$ is the Fourier transform. This calculation shows that  $I_\varLambda$ is the Fourier transform of 
$$\gamma_\varLambda = \lim_{n\to \infty} \frac{1}{|B_n|}  \sum_{x,y \in \varLambda\cap B_n}   \delta_{x-y}.$$
This object $\gamma_\varLambda$ is known as autocorrelation or  Patterson function (though it is a measure).  

In a theoretical sense the investigation of diffraction is thus reduced to the
following two step procedure: (1) Calculate the Patterson function. (2) Take
its Fourier transform.

While it is possible to  carry out this procedure for various
examples, it is far from obvious how to do this in the general case. For this
reason a lot of effort has been put into finding general criteria for the
investigation of point spectrum.   In this context, the basic questions  are the
following:

\begin{itemize}
%\item When is $\I_\varLambda$ a  pure point measure?
\item When is $I_\varLambda$ a pure point measure?
\item Where are the Bragg peaks?
\item What are the intensities of the Bragg peaks?
\end{itemize}

Some answers to these questions will be discussed in subsequent sections.

\section{Pure point diffraction}\label{Pure}
On the conceptual level two approaches to pure point diffraction have been developed in the last ten or so  years. These are given, respectively, by
\begin{itemize}
\item considering the associated dynamical systems, or 
\item investigating almost periodicity properties. 
\end{itemize}
In this section we will be concerned with  these two approaches. 

\smallskip

\subsection{Pure point diffraction via dynamical systems }
When dealing with disordered systems in statistical mechanics it is quite
standard not to consider a single object but a whole ensemble of objects
exhibiting the "same type of disorder".  In the range of aperiodic order the
same reasoning can be applied. It suggests to consider not a single set
$\varLambda$ but rather the ensemble of all subsets $\varGamma$ of Euclidean
space which have the same local structure as $\varLambda$.  This ensemble will
be denoted by $\varOmega$.  It can be made into a compact topological space
\cite{RW,Sol,LP} but we will not worry about this here.  Instead we will note
the following crucial feature of $\varOmega$: If $\varGamma$ belongs to
$\varOmega$ then so does its translate by $t\in \RR^N$ given as
$$
t + \varGamma =\{t +x : x\in \varGamma\}.$$
The obvious reason is that $\varGamma$
and $t +\varGamma$  have the same local structure.  Thus, $\varOmega$
together with the translations gives a dynamical system. We will write
$(\varOmega,\RR^N)$ to denote this structure.

In many situations $\varOmega$ comes with a canonical translation invariant
measure $m$. In these cases it is possible to provide an autocorrelation
$\gamma=\gamma_m$ by a closed formula invoking only $m$ and not using a limit.
This was first realized by Gou\'{e}r\'{e} \cite{Gou1}.  While his further
considerations use the theory of stochastic processes and Palm measures, the
formula for $\gamma_m$ can rather directly be given. Following Baake/Lenz
\cite{BL1} we obtain for the application of the measure $\gamma_m$ to the
continuous compactly supported function $\varphi$
$$
\gamma_m (\varphi) = \int_\varOmega \sum_{x,y\in \varGamma} \varphi (x)
\sigma (x-y) d m (\varGamma),$$
where $\sigma$ is a function on $\RR^N$ with
$\int \sigma (t) dt =1$.  In this way the autocorrelation is defined without a
limit. It can then be shown that it actually agrees with the corresponding
limit almost surely \cite{Gou1,BL1}.

Whenever $\varOmega$ is equipped with an invariant measure $m$ it is further
possible to consider the associated space $L^2 (\varOmega,m)$ of square
integrable functions on $\varOmega$. The translations on $\varOmega$ induce a
unitary representation of $\RR^N$ on $L^2 (\varOmega,m)$ via
$$ (T_t f)(\varGamma)= f (t +\varGamma)$$
for $t\in \RR^N$. A function $f\in L^2 (\varOmega,m)$ is called an eigenfunction to the eigenvalue $\xi$ if 
$$
T_t f =\exp (i t\xi) f$$
for all $t\in \RR^N$. If there exists a basis of
$L^2 (\varOmega,m)$ consisting of eigenfunctions, the dynamical system
$(\varOmega,\RR^N)$ is said to have pure point spectrum.  The main result on
pure point diffraction in the context of dynamical systems now reads as
follows.

\begin{theorem}
Let $(\varOmega,\RR^N)$ together  with an invariant measure $m$ be given. Then, the following assertions are equivalent:

\begin{itemize}
\item[(i)] The Fourier transform $I_m$ of $\gamma_m$ is a pure point measure. 
\item[(ii)] The dynamical system $(\varOmega,\RR^N)$ with the measure $m$ has pure point spectrum. 
\end{itemize}
In this case the group generated by the Bragg peaks $\langle \{ \xi : I_m (\{\xi\}) >0\} \rangle$ is the group of eigenvalues of the dynamical system. 
\end{theorem}

This theorem gives a characterisation of pure point diffraction in terms of
the dynamical system. Moreover, it provides further information on the
position of the Bragg peaks.  The implication $(ii) \Longrightarrow (i)$ of
the theorem established in \cite{Dwo} has been a basic tool in establishing
pure point diffraction for concrete  classes of models \cite{Hof1,Sch, Sol, Rob2}.  The
converse implication $(i)\Longrightarrow (ii)$ allows one to set up a
perturbation theory for pure point diffraction in the context of dynamical
systems. This has been carried out in  \cite{BL2} (see \cite{CS2} for
related material as well). A particular application is the study of deformed
model sets, as done earlier in \cite{BD,Gou2}.

The theorem is the outcome of cumulated efforts over many years: The analogue
result for one dimensional subshifts was proven by Queff\'{e}l\'{e}c in
\cite{Que}. For the dynamical systems we consider here, the implication
$(ii)\Longrightarrow (i)$ is due to Dworkin \cite{Dwo} with later
modifications by Schlottmann \cite{Sch} and Hof \cite{Hof2} (see \cite{EM} for
related material as well). The full equivalence was proven for systems
satisfying a certain regularity assumption known as finite local complexity by
Lee/Moody/Solomyak in \cite{LMS}. For the systems we consider (and even more
general ones) the full equivalence was then shown by Gou\'{e}r\'{e}
\cite{Gou2}. It is possible to leave the class of point processes and to go to
measures instead as discussed in Baake/Lenz \cite{BL1} and then also in
Lenz/Strungaru \cite{LS}. The statement on the eigenvalues is implicit in the
proof of \cite{LMS}. An explicit formulation can be found in \cite{BL1}.

The theorem rises the question whether the spectrum of the dynamical system is
always given by the diffraction spectrum.  This turns out to be wrong as was
shown by van Enter/Mi\c{e}kisz \cite{EM}.  More precisely, the dynamical
spectrum always contains the diffraction spectrum \cite{Dwo}. However, it
maybe strictly richer than the diffraction spectrum as shown by examples
\cite{EM}.

The theorem does not answer the question on the intensities of the Bragg peaks. The basic physical intuition concerning this issue  is that the intensity of the Bragg peak at $\xi$  should be given by
\begin{equation}\label{BTeq}
I_\varLambda(\{\xi\}) =\lim_{n\to \infty} \left| \frac{1}{|B_n|} \sum_{x\in \varLambda\cap B_n} \exp(-ix \xi)\right|^2.
\end{equation}
Validity of this equation is sometimes discussed under the header of
Bombieri/Taylor conjecture. In fact Bombieri/Taylor use validity of this
equation in their work \cite{BT,BT2} without any further justification. By now
validity has been established by direct arguments for primitive substitution
systems by G\"ahler/Klitzing \cite{GK} and for regular models sets
\cite{Hof1,Sch}. In fact, these results follow from a conceptual approach to
\eqref{BTeq} via continuity of eigenfunctions and uniform Wiener/Wintner type
results. This has been developed by Lenz \cite{Len}.  The conceptual approach
itself is hinted at in \cite{Hof2,Lag2}.  The necessary continuity of
eigenfunctions is proven for model sets in \cite{Hof2,Sch} and for primitive
substitutions by Solomyak in \cite{Sol4}.  The corresponding Wiener/Wintner
type results had already been studied by Robinson \cite{Rob}.

\subsection{Pure point diffraction and almost periodicity}
In some way or other notions of almost periodicity have been around in the
study of pure point diffraction for quite a while. In particular, there is
work of Solomyak \cite{So} providing a connection and a discussion of Lagarias
\cite{Lag2} asking for connections. There is even a characterisation of pure
point diffraction in terms of almost periodicity in the work of
Queff\'{e}l\'{e}c in \cite{Que} for the (somewhat different) situation of
symbolic dynamics. Furthermore, Meyer's work on what is now known as Meyer sets
\cite{Mey} and subsequent discussions, see e.g.  \cite{Moo1}, have a very
almost periodic flavour.  Still it seems fair to say that only with the rather
recent work of Baake/Moody \cite{BM} and Gou\'{e}r\'{e} \cite{Gou2} the
strength of this connection became apparent.

\medskip

A continuous function $f$ on $\RR^N$ is called almost-periodic if for any $\varepsilon>0$ the set of its $\varepsilon$-almost-periods
$$
\{t \in \RR^N : \| f (\cdot -t) - f\|_\infty \leq \varepsilon\}$$
is relatively dense in
$\RR^N$. Here, $\|\cdot\|_\infty$ denotes the supremum norm. Similar notions
for measures exist and give the concept of strongly-almost-periodic measure
and norm-almost-periodic measure.   
While we do not want to concern the reader
with the technical definitions here, we would like to note that norm almost
periodicity is substantially stronger than strong almost periodicity.

It is also possible to develop a concept of almost periodicity for sets. This
is done under the name of Bohr/Besicovitch almost periodicity by
Gou\'{e}r\'{e} \cite{Gou2} (see corresponding questions of Lagarias in
\cite{Lag2} as well).

\begin{theorem} Let $\varLambda$ be a Delone set whose autocorrelation $\gamma_\varLambda$ exists. Then, the following assertions are equivalent:

\begin{itemize}
\item[(i)] The Fourier transform $I_\varLambda $ of $\gamma_\varLambda$ is a
  pure point measure.
\item[(ii)] $\gamma_\varLambda$ is a strongly-almost-periodic measure.
\item[(iii)] $\varLambda$ is Bohr/Besicovich almost periodic. 
\end{itemize}
\end{theorem}
For sets $\varLambda$ whose autocorrelation is supported on a uniformly discrete set, the result is proven by Baake/Moody \cite{BM}. In the general form stated above this theorem is due to Gou\'{e}r\'{e} \cite{Gou2}. 

As will be discussed in the next section, one is particularly interested in the case of Meyer sets $\varLambda$. These are Delone sets with the property that   $\varLambda -\varLambda$ is uniformly discrete. For such sets it makes sense to define an $\varepsilon$-almost period  or better a statistical $\varepsilon$-almost period as a $t\in \RR^N$ with 
$$\limsup_{n \to \infty} \frac{ \sharp (\varLambda \setminus (\varLambda  + t) \cup (\varLambda +t) \setminus \varLambda))\cap B_n }{|B_n|} \leq \varepsilon.$$
Here, $\sharp$ denotes cardinality.  For Meyer sets  Baake/Moody \cite{BM} have the following result. 

\begin{theorem} \label{Char_Meyer} Let $\varLambda$ be Meyer with autocorrelation $\gamma_\varLambda$. 
Then, the following assertions are equivalent:
\begin{itemize}
\item[(i)] The Fourier transform $I_\varLambda$ of $\gamma_\varLambda$ is a pure point measure.
\item[(ii)] $\gamma_\varLambda$ is a norm-almost-periodic measure.
\item[(iii)] For any $\varepsilon >0$ the set of statistical  almost-$\epsilon$-periods of $\varLambda$ is  relatively dense.  
\end{itemize}
\end{theorem} 
Let us emphasise that the actual setting  and results of  \cite{BM} are considerably more general than  discussed in this theorem.  In particular, \cite{BM}  deals with weighted point sets on locally compact abelian groups. Moreover, it gives a natural connection between pure point diffraction and cut and project scheme (see below for details).

\section{Cut and project schemes and model sets} \label{Cut}
The two most prominent classes of mathematical models for aperiodic order are
models coming from primitive substitutions and models coming from cut and
project schemes.  The latter are often discussed under the name of model sets (see e.g. \cite{Moo1,Moo2} for further discussion and references).
They provide also standard examples discussed in the physical literature.
There is a wealth of results on model sets and cut and project schemes. Here,
we focus on the following issues:

\begin{itemize}
\item explicit computation of $I_\varLambda$ for regular model sets, carried out by Hof
  \cite{Hof1,Hof2} (see later generalisations \cite{Sch,BM} as well),
\item existence of a lot of point diffraction for general sets associated to cut and project schemes, shown by  Strungaru \cite{Str},
\item a natural connection between cut and project schemes and pure point diffraction, discovered by Baake/Moody \cite{BM}, and then further explored in \cite{LM,MS,BLM},
\item characterisation of primitive substitutional sets with pure point
  diffraction as model sets, due to Lee \cite{Lee} (see the work of Barge/Kwapisz \cite{BK,Kwa} for  an analogous one dimensional result in a slightly different context).

\end{itemize}

The first result justifies mathematically the corresponding parts of the
physical literature. The second result shows that order in the sense of Meyer
condition implies order in the sense of a large point component in the
diffraction spectrum. The third result (or rather the corresponding circle of
ideas) shows that cut and project sets arise very naturally within the
framework of pure point diffraction for Meyer type sets.  The final result
shows that the "other class of examples" viz primitive substitutions is not
really a different class when it comes to models with pure point diffraction.

\medskip

Let us now start by shortly recalling the framework of a cut and project
scheme.  Besides the physical space $\RR^N$ a cut and project scheme has two
further ingredients. These are a further space and a lattice. The further
space is known as internal space, perpendicular space or reciprocal space. It
will be denoted by $H$. It does not need to be an Euclidean space. It suffices
if it is a locally compact Abelian group. The lattice is denoted by
$\widetilde{L}$. It is a lattice in $\RR^N \times H$. Its projections to the
physical space and the internal space will be denoted by $L$ and $L^\star$
respectively. A precise definition of cut and project scheme now runs as
follows.

A cut and project scheme over $\RR^N$ consists of a locally compact Abelian group $H$,
 and a lattice $\widetilde{L}$ in
$\RR^N\times H$ such that the canonical projection $\pi : \RR^N\times H
\longrightarrow \RR^N$ is one-to-one between $\tilde{L}$ and
$L:=\pi(\widetilde{L})$ and the image $\pi_{\rm
int}(\widetilde{L})$ of the canonical projection $\pi_{\rm int} :
\RR^N\times H\longrightarrow H$ is dense. Given these properties of
the projections $\pi$ and $\pi^{}_{\rm int}$, one can define the
$\star$-map $(.)^\star\!: L \longrightarrow H$ via $x^\star :=
\big( \pi^{}_{\rm int} \circ (\pi|_L)^{-1}\big) (x)$, where
$(\pi|_L)^{-1} (x) = \pi^{-1}(x)\cap\tilde{L}$, for all $x\in L$.
We summarise the features of a cut and project scheme in the
following diagram:
\begin{equation*} \label{candp}
\begin{array}{cccccl}
    \RR^N & \xleftarrow{\,\;\;\pi\;\;\,} & \RR^N\times H &
        \xrightarrow{\;\pi^{}_{\rm int}\;} & H & \\
   \cup & & \cup & & \cup & \hspace*{-2ex}  \\
    L & \xleftarrow{\; 1-1 \;} & \tilde{L} &
        \xrightarrow{\,\; \mbox{\small dense}\;\,} & L^\star & \\
   {\scriptstyle \parallel} & & & & {\scriptstyle \parallel} \\
    L & & \hspace*{-38pt}
    \xrightarrow{\hspace*{47pt}\star\hspace*{47pt}}
    \hspace*{-38pt}& & L^\star
\end{array}
\end{equation*}
We will assume that the Haar measures on $\RR^N$ and on $H$ are chosen in
such a way that a fundamental domain of $\tilde{L}$ has measure $1$.

Given a cut and project
scheme, we can associate to any $W\subset H$, called the window or atomic surface, the
set
\begin{equation*}
   \oplam(W) \; := \; \{ x\in L : x^\star \in W \}
\end{equation*}

A set of the form $t+ \oplam(W)$ is called model set if the  window $W$ is relatively compact with
nonempty interior.  

\begin{theorem} \label{Meyer} Let $\varLambda$ be a Delone set. Then, the following assertions are equivalent:
\begin{itemize}

\item[(i)] $\varLambda -\varLambda$ is uniformly discrete. 

\item[(ii)] $\varLambda$ is a subset of a model set. 

\item[(iii)] There exists a finite set $F$  with
$ \vL -\vL \subset \vL + F.$
\end{itemize}
\end{theorem}
The equivalence of (i) and (ii) is due to Meyer \cite{Mey} and Moody
\cite{Moo1}. The equivalence of (i) and (iii) is due to Lagarias \cite{Lag}.
The sets characterised in the previous theorem are known as Meyer sets.  Note
that all three conditions appearing in the theorem can be understood as
indicating long range order in form of  a weak lattice
type condition.  Various further characterisations can be found in the
literature \cite{Moo1}.

\subsection{An explicit  formula for $I_\varLambda$ }

Let a cut and project scheme $(\RR^N,H, \widetilde{L})$ be given. Let also a
sufficiently nice window $W$ in $H$ be given. Sufficiently nice means roughly
speaking that the window is not a fractal. More precisely, we require $W$ to
be compact with non empty interior and and a boundary of measure zero.  In
this case, one can calculate explicitely the diffraction measure $I_\varLambda
$ \cite{Hof1,Sch}.  We need the dual lattice $\Lperp$ of $\widetilde{L}$ given
by
$$\Lperp :=\{ (k,u) \in\widehat{\RR^N}\times \widehat{H} : e^{i k l }
u(l^\star) =1\;\:\;\mbox{for all $(l,l^\star) \in \widetilde{L}$}\}.$$
Let
$\Lnull$ be the set of all $k\in \RR^N$ for which there exists $u\in
\widehat{H}$ with $(k,u)\in \Lperp$.  This set is sometimes known as
reciprocal lattice. It can be shown that there exists a unique group
homomorphism $\star : \Lnull\longrightarrow \widehat{H}$ such that
$$\tau: \Lnull \longrightarrow \Lperp, \,\:\; k\mapsto (k,k^\star)$$
is
bijective.  Then, the diffraction measure $I_\varLambda$ is given by
$$
\gammahat = \sum_{k\in \Lnull} A_k \delta_k,\:\;\mbox{with}\;\: A_k =
|\int_W (k^\star,y)dy |^2.$$
If $H$ happens to be an Euclidean space as well, the formula for $A_k$ reads $|\int_W \exp(i k^\star y) dy |^2$.  

\subsection{A lot of point diffraction for Meyer sets}
In this subsection we highlight the following result of Strungaru \cite{Str}.
\begin{theorem} Let $\vL$ be  Meyer with autocorrelation
  $\gamma_\vL$. Then, $\gammahat$ has a relative dense set of Bragg peaks.
\end{theorem}
As mentioned already this  result can be understood as linking  two notions of long range order.  The result
 is rather general as it does not assume any further regularity properties of the point set in question.

\subsection{A natural cut and project scheme}
It is a fundamental insight of Baake/Moody \cite{BM} that any set with a sufficiently almost periodic autocorrelation comes with a natural cut and project scheme. The required  almost periodicity of the autocorrelation in turn is equivalent to  pure point diffraction whenever the autocorrelation is supported on a uniformly discrete set.   
This ties pure point diffraction and cut and project schemes (within the context of  Meyer type sets).

A crucial step in the argument of \cite{BM} is to use the autocorrelation
function to introduce a topology on point sets. Taking completions with
respect to this topology then yields the internal space. In this way, cut and
project schemes lie at the crossroads of two topologies: the local topology
and the topology coming from the autocorrelation function. This point of view
is further developed in \cite{LM,Moo4,MS}.

These results allow one to derive the characterisation of pure point
diffractiveness for Meyer sets given in Theorem \ref{Char_Meyer} above. They
can also be used to characterise the dynamical systems arising from regular
model sets. This has been discussed by Baake/Lenz/Moody \cite{BLM}. As
\cite{BM} provides a cut and project scheme, the main task in \cite{BLM} is to
construct and study the window using properties of the dynamical system. The
basic connection between the dynamical system and the cut and project scheme
is given by the so called torus parametrisation \cite{BHP,Sch}.  The torus
parametrisation turns out to be strongly  determined by properties of the
eigenfunctions.  Two properties of the eigenfunctions  are central.
These are continuity of eigenfunctions and their separation properties.

As a byproduct one obtains a characterisation of lattices within Meyer sets
with pure point diffraction.  The work \cite{BLM} also plays an important role
in the investigation of substitution systems discussed next.

\subsection{Substitutions with pure point diffraction }
As mentioned already the most important classes of examples for aperiodic order are  model sets and sets arising from primitive substitution.  By their very construction models arising from primitive substitutions have a strong form of self-similarity. 

Here, we will shortly discuss a remarkable result of Lee
\cite{Lee} (see \cite{BK}  as well) relating   primitive substitution sets to model sets  within the context of pure point diffraction. In some sense the result shows that one can not get away from model sets when dealing with pure point diffraction.  
More precisely, Theorem 5.5 of \cite{Lee} gives in particular the following. 
\begin{theorem} Let $\varLambda$ be induced by a primitive substitution with finite local complexity. Then, the following assertions are equivalent:
\begin{itemize}
\item[(i)] $\varLambda$ has pure point dynamical spectrum.  
\item[(ii)] $\varLambda$ is an inter model set. 
\end{itemize}
\end{theorem}
Let us point out that the notion of a set associated to a primitive substitution requires some care.   One possibility  is to consider tilings arising as fixed points of a primitive substitution. Then a point is chosen  in each tile in a consistent way (i.e. so that points  for tiles of the same type are in the same relative position in the tile).  Thinking of the points in different classes of tiles as marked with different colours we obtain a coloured set. Such a set is underlying the statement of the previous theorem.  We refrain from giving precise definitions for a coloured set   but continue to further  explain the statement of the theorem.  

The condition (i) is  equivalent to a (suitably defined) notion of pure point diffraction for coloured sets. The condition (ii) also has to be understood for coloured sets. By definition an inter model set agrees with a model set up to points induced by the boundary of the window.  Instead of using tilings one can directly deal with substitutions for  point sets, see Theorem $5.3$ in \cite{Lee}. In this case the  additional assumption of legality of certain clusters has to be imposed.

The proof of these results winds together  different strings of reasoning.
One such string  concerns so called coincidence conditions. They give criteria for a
primitive substitution to have pure point diffraction. Building up on earlier
work  \cite{LM2,LMS2} Lee gives a new coincidence condition allowing for a characterisation of
pure point diffraction for primitive substitutions.  This characterisation is in fact part of the main result.  It is  done under the
assumption that the set in question is Meyer. The second string then is a
result of Lee/Solomyak \cite{LSo} showing that a primitive substitution with
pure point diffraction must be Meyer. As a third ingredient  Lee uses the recent results of
Baake/Lenz/Moody \cite{BLM} providing a characterisation of regular model sets
in terms of the associated dynamical systems.

\section{Open questions}
In this section we present various issues and questions for  further research. 

\begin{itemize}

\item Geometric implications of  pure point diffraction.

\end{itemize}

The phenomenon of pure point diffraction seems to be fairly well understood
within the context of Meyer sets.  This poses the question whether pure point
diffraction in itself together with mild geometric restrictions (as e.g.
finite local complexity and repetitivity) forces the Meyer property.  For
primitive substitutions this has been answered affirmative by Lee/Solomyak
\cite{LSo}. The general case seems to be open.  A particular instance of this
type of issue is the question which properties single out the lattices within
the sets with pure point diffraction.  For further discussion of these and
related issues we refer to the article of Lagarias \cite{Lag2}.

A further issue in this context is the validity of Bombieri/Taylor conjecture
(discussed above) for larger classes of examples.

Another question concerns entropy.  Of course, entropy should somehow vanish
for models with aperiodic order.  Indeed, sets with pure point diffraction and
further regularity can be shown to have vanishing topological entropy
\cite{BLR}. On the other hand,  there are natural examples of sets with
pure point diffraction exhibiting positive topological entropy. Such an
example is given by the set of visible points as shown by  Pleasants\cite{P}. It
seems that this is related to  cut and project schemes  with a
window with a "thick" boundary.

\begin{itemize}
\item  Mixed spectra and random systems.
\end{itemize}

The understanding of mixed spectra is very much at the beginning. Let us
illustrate this by considering two extreme cases:
On the one hand there are primitive substitutions models. These models carry a
lot of order by their very construction. Still, not all primitive
substitutions have point spectrum, let alone pure point spectrum. Thus, mixed
spectra go well with a very rigid order structure.  This is a conceptual issue
in the understanding of order as encoded by spectral properties.
On the other hand there are random systems.  Random systems based on lattices
exhibit a tendency to have mixed spectra with a pure point component due to
the lattice and an absolutely continuous component due to the randomness.
While this is well confirmed in examples \cite{BH1,BS} a general treatment is
not available yet. 

Actually, random systems and substitutions are not that far apart in terms of
diffraction. More precisely, as discussed by Hoeffe/Baake \cite{BH1} it is
possible to have a primitive substitution system with the same diffraction as
a random system.

One reason that  diffuse spectra  are not as well understood as
point spectra is that there does not  seem to be a
good dynamical interpretation.

%\cite{BFG,BH1,BS,MPS,R} 

\begin{itemize}
\item Homometry and inverse problem.
\end{itemize}

The above considerations have been concerned with the direct problem i.e. to
determine the diffraction given the solid. Of course, the real problem is the
inverse problem. In mathematical terms this amounts to describing all
configurations leading to the same diffraction.  This is known as homometry
problem. In this context one may ask for properties shared by all solutions to
the inverse problem as well as for further restrictions making the solution
unique.

\medskip

\textbf{Acknowledgements.} It is my pleasure to acknowledge 
helpful discussions with Michael Baake when preparing the talk at
"Quasicrystals - The Silver Jubilee", Tel Aviv, 2007,  out of which this article arose. 
 I take this opportunity to thank the organisers for the invitation to a
highly  enjoyable conference.  Financial support from German
Science Foundation (DFG) is gratefully acknowledged.

%\bibitem{AG}F.~ Axel and D.~Gratias (eds),  Beyond quasicrystals (Les Houches, 1994), 
%Springer, Berlin, (1995).

%\bibitem{Pat}
%J.~Patera (ed.),
%\textit{Quasicrystals and Discrete Geometry}, 
%Fields Institute Monographs, vol.\ 10,
%AMS, Providence, RI (1998).

\end{document}